\documentstyle[preprint,aps]{revtex}
\newcommand{\beq}{\begin{equation}}

\newcommand{\eeq}{\end{equation}}
\newcommand{\beqr}{\begin{eqnarray}}
\newcommand{\eeqr}{\end{eqnarray}}

\newcommand{\bma}{\left[ \begin{array}{cccc}}
\newcommand{\ema}{ \end{array} \right] }
\begin{document}
\title
{ Novel Phases of  Planar Fermionic Systems.}
\author{\large Ganpathy Murthy$^{1} $ and  R.Shankar$^{2}$}
\address{$^{1}$Departments of Physics
Boston University, Boston MA 02215\\
$^{2}$Yale University, New Haven, CT 06511\\} \date{\today}
\maketitle
\begin{abstract}
We discuss a {\em  family} of planar  (two-dimensional) systems

 with the following phase strucure:  a Fermi liquid, which goes by

a second ordertransition (with non classical exponent even in

mean-field) to an intermediate, inhomogeneous state (with

nonstandard ordering momentum) , which in turn goes by a first

order transition to a state with canonical  order parameter. We

analyze  two examples:  (i) a  superconductor in a parallel

magnetic field (which was  discussed independently by

Bulaevskii)for which the inhomogeneous state is obtained

  for $1.86 T_c \stackrel{\sim}{<} B \stackrel{\sim}{<} 1.86 \sqrt{2}
  T_c$ where $T_c$ is the critical temperature (in Kelvin)
of the  superconductor without a field and $B$ is measured in

Tesla, and (ii) spinless (or, as is explained,  spin polarized)

fermions near half-filling where a similar, sizeable window (which

grows in size with anisotropy) exists for the intermediate CDW

phase at an ordering  momentum different from $(\pi , \pi )$. We

discuss the experimental conditions  for realizing and observing

these phases   and the Renormalization Group approach to the

transitions.
\end{abstract}
\vspace{.75in}
murthy@buphy.bu.edu\\
shankar@cmphys.eng.yale.edu
\newpage
\section{Introduction}

Zero temperature quantum critical phenomena have  been a subject of
intense
theoretical and experimental investigation in the recent years.
Of particular interest to us has been  the Renormalization Group (RG)
description of the transitions. In our pursuit of  the RG
description of two-dimensional fermionic systems, we came across a
{\em
  family} of systems which exhibit similar phase structure and
similar
phase transitions separating the phases in mean-field theory.  Since
this phase structure seems generic and experimentally accessible
today, we have
chosen to make it  the focus of this paper,   only briefly discussing
the RG
program that led us to them in the first place.

The systems in question exhibit three phases as a control parameter
$I$
is varied.  Since the physics depends on the ratio $W/I$,  where $W$
is the
interaction strength, we can either vary $I$ at
fixed $ W $ (which is what happens in experiment), or  vary $W$
at fixed $I$  in a theoretical discussion.  The phases of the system
are
a weak coupling (large $I$) Fermi liquid phase, and a strong coupling
phase with conventional order, separated by an intermediate,
inhomogeneous phase with nonstandard ordering momentum.  The
intermediate
phase, which is the most
interesting, is {\em accessible over a broad window in $I$} (in clean
systems)
thanks to two-dimensionality. Indeed, in anisotropic systems that are
more one-dimensional, the window is broader.

We will illustrate our theme using two concrete examples: a system of
electrons with an attractive coupling between up and down spins and a
system of spinless fermions with nearest neighbor repulsion. (We
hasten
to add that we will indicate how these ``spinless fermions" are to be
realized experimentally by polarizing ordinary electrons in a manner
to
be specified.)

Consider first the system of free electrons -- by which we always
mean
Landau's
quasiparticles -- with a circular Fermi surface (FS).  They are
unstable
to the BCS state at arbitrarily small attraction $ W $. The
instability
may be traced back to the time-reversal symmetry of the spectrum and
hence of
the FS:   $E( \vec{K})= E(-\vec{K})$.

Consider next a system of free spinless fermions, at half-filling,
with
nearest neighbor hopping on a rectangular lattice. This system is
unstable to arbitrarily weak nearest neighbor repulsion $W$ which
drives
it to a Charge Density Wave (CDW) state at momentum $\vec{Q} = (\pi ,
\pi )$.  The immediate instability is again due to a special symmetry
of
the FS: due to particle-hole symmetry and nesting, if $\vec{K}$ lies
on
the Fermi surface, so does $\vec{K}+\vec{Q}$.

Both

 instabilities have  a nice RG
description. It was
shown in Reference\cite{RMP} that in an RG scheme in which modes
around the Fermi surface are symmetrically eliminated in thin slices,
the Fermi
liquid
appears as a fixed point and the above mentioned interactions appears
as a
marginally relevant terms. The existence of a steady flow of the
coupling in the
simple-minded
mode elimination scheme was also seen to result from the special
symmetry of
the Fermi surface.

We started by asking if such an RG description could be extended to
transitions
at
non-zero coupling .  A necessary prelude to such an
enterprise was finding systems which exhibited  such transitions. It
was at
that stage that we ran into the systems discussed here.

Consider first the superconductor. To move the transition to a
nonzero, but  small coupling,   (which   will
allow a perturbative RG analysis ),  we must destroy the
time-reversal symmetry of its FS by a  small  amount.  To this end,
let us
apply a magnetic field $B$
{\em parallel} to the plane. This parallel field has no effect on the
orbital motion but causes a Zeeman splitting of size $ 2 \mu B \equiv
2I$, where $\mu$ is the electron's magnetic moment. We expect that
at very small fields,  the superconductor will be stable, while for
very large
fields the Fermi  liquid will be stable.

Our strategy for showing that there  exists  an  intermediate
inhomogeneous
phase  is as follows. We start at the BCS end and   ask when the BCS
state
yields to the Fermi liquid  as the field is {\em raised}. Then we go
the Fermi
liquid end at large $I$ and ask what happens to it   as the field is
{\em
lowered}. We  find that it first becomes unstable to inhomogeneous
pairing {\em
before} the BCS state beats it in energy.  There is clearly a window
when
neither the BCS nor the Fermi liquid  can be the ground state. Here
the inhomogeneous state prevails and it is to this state, and not the
Fermi liquid that the BCS state succumbs as the field is raised.

Here are some details on the implementation of our  strategy.
Consider the
response of the BCS paired state to a small magnetic field $B$.
Since
the spins form a singlet they ignore the field altogether. In the
meantime, as $B$ increases, the Fermi liquid energy decreases because
it
is getting polarized.  Beyond the
Chandrasekhar-Clogston\cite{chandclog}
limit $ I_F= \Delta_0 /\sqrt{2}$ (where $\Delta_0$ is the BCS order
parameter in the absence of the field), the Fermi liquid has lower
energy than the BCS state and one may expect a first order transition
(hence
subscript $F$ on $I_F$ ) to
the polarized Fermi liquid (PFL). But what happens is that by this
point a
state with inhomogeneous order has beaten the Fermi liquid,  and the
first
order transition is really to this state.

To see all this,  so  let us go to  the weak coupling end (large $I$)
 where it is natural to first  find the FS in the presence of the
field and
{\em then} turn on the attraction. The radii of the up and
down FS's now differ by $q = 2I/v_F$, where $v_F$ is the Fermi
velocity,
{\em hereafter set equal to unity}.  The PFL is
stable to pairing fluctuations, at least for small coupling or large
$I$.  If  we  perform  the sum over
Cooper bubbles and evaluate the pairing susceptibility {\em as $I$ is
lowered},
we find  two interesting features:

\begin{itemize}
\item  The PFL susceptibility  first diverges  at $I_q =
\Delta_0$, {\em which is $\sqrt{2}$ times bigger than  the $I_F$ at
which the
  BCS  energy   rises  above  that of   the PFL}.
\item  This   pairing is  at momentum  $q= 2I$ .
\end{itemize}
{\em Thus
we have a window $$ \Delta_0 /\sqrt{2} = I_F  < I < \Delta_0 = I_q$$
when the
BCS state has a higher energy that the PFL, which itself is unstable
to the
inhomogeneous state.}
To repeat, as $I$ is increased, the BCS state will yield, by a first
order
transition,
not to the PFL, but
to an {\em inhomogeneous state} with wave number $q \simeq 2I$.
The phase diagram is shown in Figure 1a. Actually the inhomogeneous
state will
prevail even slightly below the lower limit of  the above window,
since at the first order
transition the   BCS state has to beat, not the PFL, but  the
inhomogeneous state

which has a lower energy.   (For completeness  we
mention that at a still lower field, $I_0 = \Delta_0 /2$ the $q=0$
susceptibility of the Fermi liquid  diverges. This is however not
very
significant  since the corresponding instability is  preempted by the
other
transitions. )

Now for the spinless fermions.  Here  we destroy the perfect nesting
of the FS
 by introducing a chemical potential
$I$. (This possibility was raised in Reference \cite{RMP} along with
another option: adding a second neighbor hopping term, which we  will
consider
only briefly.)  We establish the window as before. Let us start at
strong
coupling
(small $I$), and  consider a CDW
state at half-filling, with an order parameter $\Delta_0$. Now
imagine
modifying the chemical potential. {\em The system does not respond to
a
change in the chemical potential which is moving in the gap, exactly
the
way the superconductor did not respond to the applied field.} Once
again
a mean-field calculation shows that the $(\pi , \pi ) $ CDW  and the
FL have
equal
energies at $I_F=\Delta_0/\sqrt{2}$, where $\Delta_0$ is the
condensate without
the chemical potential term. .  At the weak coupling end, the FS
has shrunk, as shown in Figure 2, and becomes stable against
particle-hole pairs condensing at $\vec {Q}=(\pi ,\pi )$ or any other
momentum.
 An
RPA calculation shows that as $I$ is lowered, the FL becomes
unstable to a CDW
at a momentum
different from $(\pi , \pi )$ at $I_p\approx r\Delta_0$, where $r>1$
is
an anisotropy parameter which measures the ratio of the hopping in
the
$y$ and $x$ directions.  {\em Thus  we have a window $ \Delta_0
  /\sqrt{2} < I \simeq r\Delta_0 $ when an inhomogeneous state must
prevail.  This
state is inhomogeneous in the sense that it is characterized by a
momentum different from the canonical value, which in this case
happens to be
$(\pi
, \pi )$.} The phase diagram is shown in Figure 1b. (The significance
of
the points $I_q$,  and $I_0$ which do not correspond to  transitions,
will be
explained in section III.)

In both the BCS and CDW  cases the mean-field second order transition
from the
FL to the
inhomogeneous state was found to have  non-classical exponents. For
example
$\beta = 2$
for the superconductor. This nonclassical answer in a mean-field
calculation may be ascribed to the fact that the mode that becomes
unstable has a momentum which is a singular point of the
susceptibility.
This singularity and the existence of a fairly large window for the
inhomogeneous phase are both a result of the two-dimensionality of
the
system.

The plan of the paper is as follows: In the next section we will
discuss
the superconductor and relate it to earlier work. In trying to track
down existing literature in the case of the superconductor, we were
invariably directed to the works of Larkin and Ovchinnikov, and Fulde
and Ferrel\cite{LOFF}, and we shall briefly discuss these.  However,
upon digging further, we found some work of Bulaevskii\cite{bul} to
be  far
more relevant, as will be discussed.  In section III  we will
perform a similar analysis of the spinless fermion system and
describe
how it may be experimentally realized. Section IV is devoted to the
status report of the RG program. We end with conclusions and avenues
for
future research in section V.

\section{The planar superconductor}

We will begin with noninteracting electrons lying within an annulus
of
thickness $2\Lambda$ centered around the Fermi circle of radius
$K_F$.
We will construct a mean-field hamiltonian that pairs the up and down
electrons whose momenta lie within this band.

The energy of an up electron is

\beq
E_{+}(\vec{K})  = I + ({K^2 - K^{2}_F \over 2m}) = I + v_F k= I + k
\eeq
where the Fermi velocity $v_F$ has been set equal to unity, and
\beq
k = K - K_F.
\eeq

Terms higher order in $k$ are dropped.

The energy of its partner, a down spin electron of momentum $-\vec{K}
+
\vec{q}$, is

\beq
E_{-} ( -\vec{K} + \vec{q}) = -I + k - q \cos \theta
\eeq
where $\theta$ is the angle between $\vec{K}$ and $ \vec{q}$.

(It is understood that  $- \vec{K} + \vec{q} $ is also  constrained
to lie in
the same annulus.
This means that certain electrons will not have  partners for pairing
in a
state of
momentum $q$. They  must be handled correctly.)

A useful combination is
\beq
 {E_{+}(\vec{K}) - E_{-} ( -\vec{K} + \vec{q})\over 2} = I + (q/2)
\cos \theta
\equiv z(\theta ).
\eeq

Let us begin with the mean-field hamiltonian
\beqr
H &=& {\Delta \Delta^{*}  \over  W } \nonumber \\
& & +
\int_{-\Lambda}^{\Lambda} dk \int_{0}^{2\pi}
d\theta
 \left[ c_{+}^{\dag}(\vec{K}) c_{+}^{}(\vec{K}) (I + k) +
c_{-}^{\dag}(\vec{K}) c_{-}^{}(\vec{K}) (- I + k )  \right. \nonumber
\\
& & -  \Theta ( \Lambda - ||\vec{K} - \vec{q}| - K_F |) \left. \left[
\Delta
c^{\dag}_{+}( -\vec{K} +
\vec{q}) c^{\dag}_{-}( \vec{K} ) + \Delta^{*}  c^{}_{-}(
\vec{K} ) c^{}_{+}( -\vec{K} + \vec{q}) \right] \right]
\eeqr
where  the condensate was taken to have the form $ \Delta (x) =
\Delta \ exp \ [
i \vec{q} \cdot \vec{r} ]$ ,  the step function
$\Theta ( \Lambda - ||\vec{K} - \vec{q}| - K_F |)$ ensures that
the other partner in the pair also lies in the annulus, and the
subscripts
refer to the spin.  The reader should not worry about
factors of $2\pi$ since all key results will  be given in term of
observables.

By going to Bogolubov operators $d_{\pm}$ and performing the  usual
transformation that kill the ``bad'' terms, we obtain
\beqr
H &=& {\Delta \Delta^{*}  \over  W } \nonumber \\
&+&
\int_{-\Lambda}^{\Lambda} dk \ \int_{0}^{2\pi}
d\theta  \  \Theta ( \Lambda - ||\vec{K} - \vec{q}| -
K_F |)
   \nonumber \\
& &\left.\left[ k - I - q \cos \theta
+ d^{\dag}_{+} d_+\left[ I + (q/2)  \cos
\theta + \sqrt{\Delta^2 + (k - (q/2) \cos \theta )^2}
\right]\right]\right.
\nonumber \\
&+ &  \left.
(d^{\dag}_{-} d_{-}-1) \left[ -I - (q/2) \cos \theta + \sqrt{\Delta^2
+
(k -   (q/2) \cos \theta )^2} \right] \right. + {\pi q^2 \over 2} -
4 \Lambda q.
\eeqr
  The last two terms come from the unpaired electrons.   The ground
state is
found by
filling all negative energy states.  (If we set $\Delta = 0$ above,
the ground
state energy
should reduce to $- 2 \pi (\Lambda^2 +  I^2) $, the energy of the
PFL. ) If
we subtract off the energy of the PFL, we find the relative energy
\beqr
E(\Delta) - E(0) &=& 2 \pi I^2 + {\Delta^2 \over  W } + \pi ({q^2
\over 4}  -
\Delta^2) -  2 \pi \Delta^2 \ln {2 \Lambda \over \Delta} \nonumber \\
& & + \int_{0}^{2\pi} d\theta \left[ \theta (z - \Delta) \left[ -z
\sqrt{z^2 -\Delta^2}+ \Delta^2 \ln {z + \sqrt{z^2 - \Delta^2} \over
\Delta }\right] + z \to -z \right] \label{seven} \\
z(\theta ) &=& I + {q \over 2}  \cos \theta \nonumber
\eeqr
where $\Delta$  and $\Delta^2$ stand for $|\Delta |$ and $|\Delta
|^2$.

Most of our results are based on an analysis of this equation.

The first term comes from subtracting off the polarization energy of
the
PFL. As for the $\theta$ integral let us note that if $\Delta > I +
q/2$, it does not contribute.  It is clear that nonzero $q$ is a
liability in this region and the minimum energy  is readily found to
occur at $\Delta_0$ given by
\beq {1 \over  W } = 2 \pi \ln {2 \Lambda \over \Delta_0}
\eeq
exactly as if $B$ were never turned on. This is just the
manifestation  of
the fact that the BCS system ignores the field. The condensation
energy
(relative to the  PFL energy) is found to be
\beq
E(\Delta ) - E(0) = \pi (2I^2 - \Delta_{0}^{2})
\eeq
which leads to the result  that $I_F =
\Delta_0/\sqrt{2}$ is the field at which the BCS system yields to the
Fermi liquid. (Since $\Delta_0 = \sqrt{2} I_F > I_F$ the answer is
consistent with the assumptions made in deriving it.)

Now for the weak coupling (or large $I$)  Fermi liquid end. Either
by summing
all the Cooper bubbles, or by
 taking  the $\Delta^2$ derivative of the energy at $\Delta
=0$, we find the formulas for $\chi^{-1}$. (In taking
this derivative it is important to note that if $I > (q/2)$,
$z(\theta )$
is always positive and only the first term in the $\theta$ integral
contributes, whereas if $(q/2)>I$, the second contributes for $\Delta
=
0$.) The result is

\beqr
\chi^{-1} ((q/2)) &=& {1 \over  W } - \int_{0}^{2 \pi } d \theta \
\ln {
\Lambda \over |z( \theta ) |}     \\
&=&  {1 \over  W } - 2 \pi \ln {2 \Lambda \over I +
\sqrt{I^2 - (q/2)^2}} \hspace{1in} (q/2)<I \\
&=& {1 \over  W } - 2 \pi \ln {2 \Lambda \over (q/2)} \hspace{1in}
(q/2)>I .\\
\eeqr

{\em Note that at each value of $I$ (the applied field), the softest
mode is the one with $(q/2)=I$ and that this is a singular point of
$\chi$.   }
This mode becomes unstable when $I = I_q$,
where
\beq
{1 \over  W } = 2 \pi \ln  {2\Lambda \over I_q}
\eeq
If we compare this equation to the gap equation in the absence of
field
\beq
{1 \over  W } = 2 \pi \ln  {2\Lambda \over \Delta_0}
\eeq
 we see that
\beq
I_q = \Delta_0
\eeq
{\em Thus we have a window
\beq
 {\Delta_0 \over \sqrt{2}}  \stackrel{\sim}{<} I \stackrel{\sim}{<}
\Delta_0
\eeq
wherein the neither the Fermi liquid, nor the homogeneous
superconductor
is the ground state.} (Actually the lower limit should be lower since
the BCS state has to beat not the PFL but the inhomogeneous state.
However the condensation energy of the inhomogeneous state is quite
small and so is the change in the lower limit.)

If we put in the numbers we find that if $B$ is in Tesla and $T_c$ in
Kelvin, the new phase should be stable in the interval $1.86 T_c
\stackrel{\sim}{<} B \stackrel{\sim}{<} 1.86  \sqrt{2} T_c$. We have
used here
the free
electron $g$ factor. In practice $g$ could be much higher, and the
values  of $B$ correspondingly
lower.

Figure 3 shows a plot of the energy in Eqn.(\ref{seven}) versus
$\Delta$.
The
parameters
$\Lambda = 100, \ I= .834,  W =.031 $ are chosen so as to bring the
BCS
energy (measured relative to the PFL) to zero, i.e., we have chosen
$I=
\Delta_0 /\sqrt{2}$. Plotted on the same graph is the energy of the
inhomogeneous state at  $ (q/2)= .923$. Note that this $(q/2)$ is not
equal
to $I$. In other words, although $(q/2)=I$ becomes unstable first,
once
the order sets in, a larger $(q/2)$ does better and the graph shows
the
energy for the best $(q/2)$. If we lower $I$ a bit, the BCS state
will
dip down very quickly to below the inhomogeneous state. Thus we may
take $I= \Delta_0 /\sqrt{2}$ to be essentially the field for the
first order
transition to the inhomogeneous state. If $I$ is raised further, the
BCS
graph will move upwards and so will the inhomogeneous state's energy
at the minimum. As the minimum moves towards the origin, the optimal
$(q/2)$ will move towards the current value of $I$. Finally for  $I >
\Delta_0 $ the PFL will become stable to all pairing fluctuations.
The transition to the PFL is clearly second order in mean-field
theory.

{}From the energy formula it is easy to compute the specific heat
by finding  the  density of states,  and the
magnetization by taking the $I$ derivative. (Since $(q/2)$ and
$\Delta$
are slaved to $I$ one may ask how they are to be handled during the
derivation. In general there will be implicit and explicit changes in
$E$ to to a change of $I$, however at a point where
$\partial E/\partial q $ and $\partial E/\partial \Delta$
vanish (as they do for us) only the explicit derivative matters.)
The
result is,  in terms of the corresponding quantities for the PFL,
\beqr
{M \over M_{ PFL}} &=& {1 \over 2 \pi I} \int_{0}^{2\pi}d \theta
\sqrt{z^2 - \Delta^2}\left[ \theta(z - \Delta ) - \theta ( - z -
\Delta ) \right] \\
{C \over C_{PFL}} &=& {1 \over 2 \pi }  \int_{0}^{2\pi} d
\theta{\left[ z \theta(z - \Delta ) - z \theta ( - z - \Delta )
\right]\over \sqrt{z^2 - \Delta^2}}
\eeqr
where $\Delta$ and $(q/2)$ are at their optimal values. Putting in
the
numbers we find that   these
ratios drop from the value of unity at the PFL end, down to $.92$ and
$.79$
respectively  by the time we reach  the situation shown in Figure 3
(which is
more or less the end
of the inhomogeneous phase).

Let us next compute the exponent $\beta$ which characterizes the
onset of $\Delta$. If we could write the energy near the transition
as
\beq E = -t \Delta^2 + u \Delta^4  + \ldots
\eeq
where $t = \Delta_0 - I$, we would find as usual that $\Delta \simeq
t^{1/2}$. If however we analyze our energy function we find that

\beq
E = -t \Delta^2 + u \Delta^{5/2} + \ldots
\eeq
leading to $\beta =2$.  In other words, {\em the energy function does
not have an analytic expansion in $\Delta$  near the transition,
which in turn
can
be traced to the fact that the optimal $(q/2)$ at the transition is
a
singular point of $\chi^{-1}$. Thus even though we are doing
mean-field theory, a nonclassical exponent emerges. }

We could go on and compute the electromagnetic response functions but
do not, for the following reasons. Our analysis has shown that in the
window

$$ {\Delta_0 \over \sqrt{2}}  \stackrel{\sim}{<} I \stackrel{\sim}{<}
\Delta_0 $$
an inhomogeneous state characterized by momentum of magnitude $q $
(which is of
the
order $I$) has lower energy than either the
PFL or the BCS state. However our analysis does not choose a
particular direction for $q$. Thus, as in problems involving, say
crystallization,\cite{Brazovskii} all modes in this circle are
degenerate. In
this case the low energy physics of the condensate is very much like
that of a
Fermi system, i.e., centered around a circle and not the origin.
Recently
Hohenberg and Swift\cite{hoswift} have shown how the shell
elimination devised
in Ref. \cite{RMP} for fermions can be applied to this problem. For
the present
simply note that,  instead of the plane wave we considered so far, a
standing
wave of the same wavelength might have a lower energy. The standard
trick for deciding what exactly happens  is to expand the energy
functional in
a
power series in $\Delta (\vec{q})$, going out to fourth order. The
fourth order vertex  (which is the particle-particle bubble with two
more insertions it it) will determine which combination of modes is
best the near transition. Unfortunately we cannot do this here since
an
analytical expansion in $\Delta$ does not exist. While we may expect
that $\beta = 2$  may be insensitive  to the actual form of the order
parameter
(plane wave
versus cosine),  features like
specific heat will be very sensitive to the true ground state. For
example if $\Delta \simeq \cos qr$, bands will be formed and the
density of states can go up above the PFL value. The only thing that
seems certain is that the specific heat  will not be activated.
Likewise the
electromagnetic response and Meissner effect (for a small
perpendicular field) will depend on the knowledge of the true ground
state.
This is a problem we have not solved. All we can say is that if the
answer is a cosine, there will be an oscillation in spin density at
the
same wavelength.

We now consider the relation of our work to that of Larkin
-Ovchinnikov and
Fulde-Ferrel\cite{LOFF}\cite{maki}.  These authors considered a three
dimensional
superconductor with ferromagnetic impurities.  The spins of these
impurities coupled to the electrons {\em via the exchange
interaction}
so that they could be represented on the average by some external
field
that coupled only to spin. These authors too predict an inhomogeneous
phase.  We now list the differences between our work and theirs.
\begin{itemize}
\item These authors consider a system coupled to magnetic impurities
while we need a clean system.
\item They evade the orbital complication of the magnetic  field
(even though
they work in three dimensions)
because the field  really represents the exchange interaction with
impurities.
We evade it by considering a planar system with an external
parallel field.
\item{\em Most importantly, they expect the inhomogeneous phase over
a very narrow  window $.707 \Delta_0 \le I \le .755 \Delta_0 $
in three dimensions
whereas we expect it over a broad window $. 707 \Delta_0 \le I \le
\Delta_0 $ in two dimensions, with $I$ as the  controllable external
field. } The window will
be even wider in anisotropic systems (e.g. elliptical FS) because
they are more one
dimensional and will have roughly parallel segments of the FS  over
longer intervals.

\item Due to the fact that the optimal $q $ is a singular value of
$\chi^{-1}$ we find nonclassical exponents ($\beta =2$) in the
mean-field  level, while they find classical exponents due to the
fact that the optimal $q$, for small $\Delta$ is $q=1.2I$, which is a
nonsingular point of the three dimensional susceptibility. This
allowed Larkin and Ovchinnikov to   study mode coupling and to decide
that the
cosine
order parameter is better than the plane wave.  Till we can do a
similar thing here it may be a reasonable to assume that the same
thing happens in $d=2$ also. In this case we can expect as they do,
that the spin density will will oscillate with period $q $
and that the specific heat will be greater than for the PFL.
\end{itemize}

Consider next the work of Bulaevskii\cite{bul}. He looks at layers in
a
magnetic
field and works out critical fields for any tilt and at finite
temperature.  His formula for zero tilt at zero temperature gives the
window we quoted.  The exponent $\beta = 2$ is reflected in the $T$
dependence of his results.  Like us, he does not consider spin orbit
coupling and for this reason could not explain some of the
experiments\cite{prober} done at that time, as pointed out by  Klemm,
Luther and Beasley\cite{KLB} , who did include this effect. {\em
  It appears that  the time is ripe to see this inhomogeneous phase.
The
  conditions are that the material be as two-dimensional as possible,
  clean, have low $Z$ (to minimize spin-orbit effects) and
  have a low $T_c$ so that the required fields are are not too large.
}
It will be interesting, not only in its own right, but also as a part
of the
family of systems discussed here. Of course at finite $T$, there will
be
no ordered phase and a Kosterlitz-Thouless phase with algebraic order
will take its place\cite{halpnel}.  The phase structure will still be
visible
at small $T$ and the $T=0$ critical point will  control the
finite $T$ physics in the quantum critical
region as in other problems \cite{CHN}\cite{sachdev}.

\section{The spinless fermion system with CDW order}

We begin by reassuring our readers that what we mean by a spinless
fermion can be experimentally realized in at least one way. In the
idealized calculation we consider a system of spinless fermions that
fills up a nested Fermi sea and ask what happens as we tamper with
the
nesting  by changing the chemical potential. Assuming that all
the action is centered around the immediate vicinity of the Fermi
surface, i.e., that the interaction and the change in chemical
potential
$I$, are small compared to the bandwidth, we can duplicate the same
physics
near the Fermi surface as follows. We take regular electrons and fill
them up to, say, 45\% (instead of half) of the Brillouin zone.  Now
we
apply a magnetic field in the plane. This does nothing to the orbital
motion but splits the spin up and down Fermi surfaces. For some
choice
of field, the bigger  of the surfaces will begin to nest, while the
other would
have moved away in the opposite direction and hopefully will not do
anything
interesting. The polarized electrons with
a nested or nearly nested Fermi surface will be the spinless fermions
of
the theoretical calculation. Of course a CDW in this case will also
imply a spin density wave or SDW.

As in Reference \cite{RMP} we will consider free fermions with the
dispersion relation which comes from nearest neighbor hopping:

\beq
E(x,y) = - \cos x -  r \cos y
\eeq

where $x$ and $y$ stand for $K_x$ and $K_y$. We choose anisotropic
hopping since the generic physics we are talking about can be
obscured
by the van Hove singularities that lie on the Fermi surface when $ r
=1$. We will work with $ r > 1$. This energy relation satisfies the
condition

\beq
E(x + \pi, y +\pi ) = - E(x,y)
\eeq

and the vector $Q = (\pi , \pi )$ connects points on the two branches
of
the Fermi surface, labeled by $\alpha = \pm 1$ as shown in Figure 2.
At
half-filling the Fermi surface is defined by $E=0$ or

\beq
y = \pm \ \cos^{-1} \left[ - {\cos x \over r} \right]  = \alpha
\cos^{-1}
\left[ - {\cos x \over r} \right]\label{cosinv}.
\eeq

We  will shortly use  a variable $\varepsilon$  that measures the
energy from
the FS of the free particles. In our problem where the FS is at zero
energy,
$\varepsilon = E$.

Imagine  turning  on a chemical potential so that

\beq
E(x,y) = - \cos x -  r \cos y +I
\eeq
The Fermi surface of the free fermions will now shrink to a smaller
size
and no longer nest.
Of relevance is the combination

\beqr
{E(x+\pi+q,y+\pi+p)-E(x,y)\over 2}  &  \equiv & z(x,y,q,p) \\
&=&
I-(q/2)\sin x-r(p/2)\sin y\\
&=& I - (q/2)\sin x- (p/2)\ \alpha \sqrt{r^2 - \cos^2x}\\
&\equiv & z_{\alpha}.  \label{z}
\eeqr
 In the above formula, $\alpha = \pm 1$ is the branch index, and, as
everywhere else in the paper, terms of quadratic order or higher in
the small quantities  $I,p,q$
will be ignored.

Once again we limit ourselves to a band of energy $\Lambda$ on either
side of the free electron FS. To this end we change variables from
$(x,y)$ to $(x,\varepsilon )$, where $\varepsilon$ measures the
energy
from the FS. (See Reference \cite{RMP} for details. )  As a result

\beq
\int dx \ dy \rightarrow \int dx d\varepsilon J(r,x)
\eeq

where the Jacobian is

\beq
J( r , x) = { 1 \over  \sqrt{r^2 - \cos^2 x}}
\eeq

on the free electron FS.  We will ignore its deviation off the FS.
 It is useful to define

\beq
\overline{J} = {1 \over 2\pi} \int_{0}^{2\pi} J(x)dx.
\eeq

The mean-field hamiltonian is

\beqr
H &=& {\Delta^2 \over W} + \int_{-\Lambda}^{\Lambda} d \varepsilon
\int_{0}^{2\pi}dx J(r,x)
\left[ c^{\dag}_{+}(\varepsilon , x ) c_{+} (\varepsilon , x)  +
c^{\dag}_{-}(\varepsilon , x) ) c_{-} (\varepsilon , x)\right]
\nonumber \\
& &+
\int_{-\Lambda}^{\Lambda} d \varepsilon \int_{0}^{2\pi}dx J(r,x)
\Theta (
| \varepsilon ( x + \pi + q, y +\pi + p  ) | < \Lambda )\nonumber\\
& &\left[\Delta (c^{\dag}_{+}(\vec{K} ) c_{-} (\vec{K} + \vec{Q} +
\vec{i} q +
\vec{j} p ) +
\Delta^{*} c_{-}^{\dag} (\vec{K} + \vec{Q} + \vec{i} q + \vec{j} p
)c^{}_{+}(\vec{K} )\right]
\eeqr
where now the subscripts $\pm$ on the operators refer to the branches
of
the FS.

To exhibit the intermediate phase, let us assemble the two
ingredients:
(i) The energy difference between the Fermi liquid and the $(\pi ,
\pi)$
CDW state and (ii) The inverse Fermi liquid susceptibility (in RPA)
to
CDW formation at momentum $(x + \pi + q, y +\pi + p )$. The space
between the
zeros of these  gives a window in $I$ for the inhomogeneous state.

A standard mean-field calculation, as in the superconducting problem,
shows that for $\Delta > I$, the energy difference between the $(\pi
, \pi ) $
CDW  state
and Fermi liquid is

\beq
E(\Delta ) - E(0) = {\Delta^2 \over W} + \int_{0}^{2\pi} dx J( r , x)
[ I^2 - {\Delta^2 \over 2} - \Delta^2 \ln {2\Lambda \over \Delta } ]
\label{E}
\eeq

This energy difference  has a minimum at
\beq
\Delta_0 = 2\Lambda\exp\left[-{1\over 2\pi\overline{J} W}\right] .
\eeq
Feeding this back into the energy formula we find that the   $I$ at
which the
$(\pi , \pi )$ CDW state loses out to the Fermi liquid is

\beq
I_F = {\Delta_0 \over \sqrt{2}}.
\eeq

(Note that once again the answer is consistent with the assumption
$\Delta > I$ that went into its derivation.)

By summing the usual particle hole bubbles, the inverse
susceptibility of the
Fermi liquid
is found to be

\beq
 \chi^{-1} (q,p)= {1 \over W} - {1 \over 2} \sum_{\alpha}
\int_{0}^{2\pi} dx J(
r , x) \ln
{\Lambda\over |z_{\alpha} |} \label{chi}
\eeq

where $z_{\alpha}$ is the value of $z$ on branch $\alpha$,
as defined in Eqn.(\ref{z}). This gives

\beq
\chi^{-1} (q,p)= {1 \over W}-\int_{0}^{2\pi} dx
J( r , x) \ln {\Lambda
\over  \sqrt{(I - (q/2)\sin x)^2 - (p/2)^2(r^2 -\cos ^2 x)}}
\eeq

First note that the uniform CDW (i.e., $(\pi , \pi )$) mode becomes
singular when $I=I_0$ where

\beq
{1 \over W} = 2 \pi \overline{J} \ln {\Lambda \over I_0}
\eeq

i.e., at

\beq
I_0 = \Lambda \exp \left[ - { 1 \over 2 \pi  \overline{J} W} \right]
=
{\Delta_0 \over 2}.
\eeq

Next we consider just nonzero $q$. The instability occurs when

\beq
{1 \over W} = \int_{0}^{2\pi} J(r, x)dx \ln {\Lambda \over |I -
(q/2)\sin
x|}.\\
\eeq

Let us consider  this equation at large $r$ where the analysis is
easier.  Now
$J \simeq 1/r$  and

\beqr
{1 \over W} &=& {2 \pi \over r} \ln {2 \Lambda \over I + \sqrt{I^2 -
(q/2)^2}}
\hspace{.5in} q/2
< I \\
&=& {2 \pi \over r} \ln {2 \Lambda \over (q/2)} \hspace{.5in} q/2 >
I.
\eeqr

The optimal value is $q/2 = I$,  and we find  upon inverting the
above that the
$I$ value at
the onset of the instability is

\beq I_q  = 2\Lambda \exp \left[ - { r
  \over 2 \pi W} \right] = 2\Lambda \exp \left[ - { 1
  \over 2 \pi \overline{J} W} \right] = {\Delta_0}
\eeq

At $r$ not too large, but also not too close to unity (so that the
singularity of the Jacobian does not dominate the integral),  similar
results
hold.

So far we have found the same window as in the case of the
superconductor,
namely

\beq
{\Delta_0 \over \sqrt{2}} < I <  \Delta_0
\eeq

But the window gets wider if we look at nonzero $p$. (This is also
clear
from Figure 2.)  Now we have

\beqr
{1 \over W} &=& \int_{0}^{2\pi} J(r, x)dx \ln {\Lambda \over
\sqrt{I^2 - (p/2)^2(r^2 - \cos^2 x)}}\nonumber\\
&=& \int_{0}^{2\pi} J(r, x)dx \ln {\Lambda \over
\sqrt{I^2 - (rp/2)^2 +(p^2/8)  + (p^2/8) \cos 2x)}}
\eeqr

For large $r$, this integral is seen to be a maximum when $p = 2I/r$
where its value is $(2 \pi /r )  \ln {2\Lambda r \over I_p} $ so
that

\beq
I_p = r \Delta_0.
\eeq

The following table shows the deviation from this result for smaller
$r$.
In all these cases the optimal $p$ was found to be   $2I/r$. Note
that the
large $r$ results are not bad for $r$ as low as $1.1$.
\begin{center}
\bigskip
\begin{tabular}{|c|c|c|c|c|}
\hline
  $ r $ & 1.1 & 1.5 & 2.5 & 3.5 \\
$ {I_p \over r \Delta_0 } $ & .83 & .93 & .98 & .99 \\
\hline
\end{tabular}
\bigskip
\end{center}
The final phase diagram is as shown in Figure 1b.  The main point is
that
as the anisotropy grows, so does the window, since the system is
becoming increasingly one dimensional and the opposite faces of the
Fermi surface are becoming globally parallel.  (In the limit of
parallel
faces there exists a perfect nesting vector for any $I$.)

 We will not
consider the general case of a nonzero $q$ and $p$ since we have
already
seen a substantial window for observing the inhomogeneous phase.  We
also do not try to compute its energy as a function of $\Delta$ (as
we
succeeded in doing for the superconductor) because of the following
problem. Recall that  in the superconductor problem we coupled $K$ to
$-K +
q$. That in turn must be coupled to $- (-K+q)+q = K$, which means
only two
modes are to be coupled. In the present case we coupled $K$ to $K + (
\pi, \pi)
+ q$, which in turn must be coupled to $K+ 2q$ and so on. We did
perform a
calculation keeping just the first coupling to exhibit a variational
wave
function which could beat the Fermi liquid and the $(\pi, \pi)$ state
inside
the window.

Due to the fact that the optimal momentum $p$ at the
transition is a singular point of $\chi$, we can once again expect
nonclassical exponents. We guess that $\beta$ will be larger than the
classical
value, probably equal to $2$ once again.

\subsection{Experimental realization}
Let us now turn to the experimental realization of this system. We
start
with a system of real electrons  not too close to
  half-filling, say at 45\% filling. At zero
temperature, this will probably be a BCS superconductor. Next we
apply a
parallel magnetic field.  This is  the  problem we just studied. The
BCS state
will eventually pass via the
inhomogeneous superconducting phase to the polarized Fermi liquid as
the field
rips apart the Fermi surfaces of the up and own spins. (Here we must
choose a
material with a large $g$ factor for the electron.)  Focus on the two
Fermi
surfaces.  One shrinks further away from half-filling and hopefully
does
nothing interesting. The FS
of the growing species will now come close to nesting as the field is
raised.
{\em This species of polarized electrons constitutes the spinless
fermions of
the calculation, while  the applied field, which adds a  constant
(Zeeman)
energy per particle, will play the role of the  changing  chemical
potential.}
Note however that as the magnetic field grows and pushes the FS
closer to
nesting, the $I$ in the spinless version {\em decreases}.   At one
point the
system should enter the  inhomogeneous CDW phase, which here  means
the spin
density  also will oscillate with a momentum slightly different from
$(\pi ,
\pi )$. Eventually  the system will jump by a first order transition
to
the $(\pi , \pi )$ state. In other words, when the chemical potential
(i.e.e, applied magnetic field)

falls within a window, the system density locks at half-filling,
taking
particles from the reservoir, which in this case is the other species
of
spins.  We should see a sudden  increase  in polarization  which
should then
remain fixed as the field is increased,  since   the up and down
densities remain
locked. (Unlike in the case of an ideal reservoir which can donate
any
number of particles at one energy, there will be a cost of energy
when a
macroscopic number of fermions have to be converted from the
shrinking FS to
the growing one. This will decrease the region of stability of the
commensurate
$(\pi , \pi ) $ CDW  phase.) If the field is raised further, we are
effectively
changing the sign of $I$ and so the same picture  will appear in
reverse: a
first order transition  to the inhomogeneous state followed by a
second order
transition to the Fermi liquid.

{\em By design, the grand canonical picture applies to the above
experiment, where
the system of polarized electrons indeed is in contact with a
reservoir (of
opposite spin electrons).}  The free parameter is the chemical
potential or the
applied magnetic field, and the particle density is chosen by the
system to
minimize its energy.

 Now  the ideas discussed above also apply to other problems where
the number
density is the independent variable,  say  when a system of regular
electrons
at half-filling, (as in a Hubbard model),  is  doped.    Whereas  if
we took a
commensurate system and changed its  chemical potential, it  will
initially
ignore it and hang on its special density, here we want to forcibly
change the
number density and see what happens. We illustrate what is to be done
in such a
case  by re-expressing the above spinless fermion analysis in terms
of number
density. In other words we ask-- Suppose we took a half-filled system
of
spinless fermions and doped it, what will it do? Figure 4, which is a
schematic,  helps us analyze this. Along the $x$-axis we vary $I$,
the chemical
potential, and cover the three phases. (The CDW phase refers to the
$(\pi , \pi
)$ state). We find (in our calculation) that it has half-filling for
all $I$ up
to the first order transition. So we plot the corresponding $n$ as a
dark
horizontal line ( at $1/2$) going  up to the transition. A slight
increase in $I$ causes a
first order transition to the inhomogeneous state. It has a lower
density
$n_q$. Further increase in $I$ causes a decline in $n$  till we reach
the
second order transition to the Fermi liquid. There is no jump in $n$.
Hereafter
$n$ and $I$ are related as in a free Fermi liquid: $  n = {1 \over 2}
- I$.
(The coefficient of $I$, the  density of states, is set to unity).
Now we can
answer the question we posed. To find out what happens at any given
density,
choose a value  for  $n$, move horizontally to the dark line and come
right
down to the phase diagram on the  $I$ axis. Thus when $n=1/2$ the
system is in
the $(\pi , \pi )$ state. As $n$ is lowered, ever so slightly, the
system  is
stuck at the first order transition and there will be coexistence of
the $(\pi
, \pi ) $ CDW  and inhomogeneous states, the ratio being determined
by $n$.
When  $n$ is lowered down  to $n_q$, the system will become  all
inhomogeneous. Thereafter  we move smoothly along the dark line to
the Fermi
liquid transition and Fermi liquid phase. Thus,  although a  slight
change in
chemical potential does nothing to the $(\pi , \pi )$ state,  the
slightest
doping causes phase coexistence.

The above discussion has focused on  breaking the nesting symmetry by
changing the chemical potential. This is the choice  with the closest
analogy to the superconductor. We have also studied a case where
nesting
is destroyed by adding a second neighbor term which  causes
additional
wiggles on the FS without changing its volume. Here we found that the
window for the inhomogeneous state shrinks. In the large $r$ limit,
it goes to
zero: the second order transition at the origin in $\Delta$ space
occurs when
the
first order transition does.  It is still interesting that the
first order transition far from the origin  will be accompanied by
huge
fluctuations (at an incommensurate momentum) close to the origin.
The reason
 the
uniform state is more stable in the presence of this kind of
modification of the FS  is clear: whereas changing $I$  ruins
nesting at $(\pi , \pi )$ uniformly over the FS, the additional term
due
to second neighbor hopping vanishes at some points on the old FS.
Indeed any modification to the dispersion relation   will necessarily
be
periodic and vanish somewhere,  unless the periodic function has a
constant component, as in the case of the

 chemical potential. Only in the latter  case, when the new
free-electron FS does not intersect the old FS anywhere, do we have a
big
window for the inhomogeneous state.

A problem  worth   studying is one where both a second neighbor term
and chemical potential  destroy the nested Fermi surface, with the
latter as
the control parameter.
\section{The Renormalization Program}

As mentioned earlier, the simple minded and intuitively appealing
procedure of eliminating thin slices on either side of the FS works
in
the case of problems where the FS has special symmetries. The reason,
as
explained in detail in reference \cite{RMP}, is as follows. When we
compute a one loop graph for the $\beta$-function, the two lines in
the
loop are either particle-hole (in the CDW case) or particle-particle
(in
the BCS case).
{\em To get a nonzero contribution to the flow, it is necessary that
  both lie in the thin shells being eliminated and also obey momentum
  conservation.} In the BCS case if the overall momentum is zero,
these
lines have equal and opposite momenta.  Given time-reversal symmetry,
they have equal energies, so that if one lies in one or the other
shell
(above or below the FS) that is being eliminated, so does the other,
no
matter what the direction of the line momentum.  In the CDW case, the
momenta of the particle and hole lines differ by $Q= (\pi ,\pi )$.
Since this reverses the energy for the nested problem, again if one
lies
(anywhere) in the shell below or above the FS that is being
eliminated,
 the other lies in the shell of
opposite energy, also  being eliminated.

In both cases, no matter how many shells we eliminate, the couplings
keep flowing, with each factor of $s$ reduction in cut-off giving a
contribution proportional to $\ln s$. The flow is off to the
condensed
state.

All this changes if we break the symmetry. Consider the CDW example.
Now in
order for one of the momenta to lie in a shell of thickness $d
\Lambda$
and scatter into another one also being eliminated, its momentum must
have a very specific {\em direction} that lies within a narrow range
of
size equal to the shell thickness divided by bandwidth. Thus in the
shell elimination game, there will be no flow. This is not an
insurmountable problem.  In reference \cite {RMP} another scheme
called
the Field Theory Scheme (FTS) is invoked.  Here one computes some
physical quantity in the cut-off theory and sets the derivative with
respect to the cut-off to zero to obtain the $\beta$-function. We
get the same flow as before for $\Lambda >>I$. The flow does change
character as $\Lambda$ approaches $I$, as it should, and indicates a
transition. But we are not very happy with this method since the FTS
is
not generally  to be used when any of the other energies in the
problem
comes close to $\Lambda$. This is because in the FTS one tries to get
away with just the quartic interaction and this requires that the
cut-off dominate all other energies.  In other words the smallness of
ratios
like $p/\Lambda$ or $q/\Lambda$ is what allows one to neglect higher
operators.
In the full fledged Wilson-Kadanoff
scheme there is, of course,  no such restriction, {\em provided one
works with
operators
  of arbitrary complexity}, a prospect we do not want to entertain.
We are
currently engaged in solving this
problem.

Another approach is to follow Hertz\cite{hertz}.  Here one couples
the
fermions to a bosonic variable $\Delta$ as in a Hubbard-Stratanovich
transformation, integrates the fermions out and works with the
effective
action for $\Delta$. As mentioned in Reference \cite {RMP}, this is
generally going to cause trouble since integrating out gapless
degrees
of freedom (anathema to the standard RG) can and typically will,
lead to a singular action for the remaining fields. If
one did this for the usual BCS problem, one finds that the action
cannot
be expanded in powers of $\Delta$.  Even with the $I$ term present,
the
action is non polynomial in $\Delta$ since the chosen momentum at the
transition to the inhomogeneous phase is a singular point of the
susceptibility. (In three dimensions this is not so because of phase
space. In one dimension the singularity is even more pronounced.)
Even
if the action is polynomial, it may not have an analytic expansion in
$\omega$ or $q$. Hertz has argued that in some cases we can still
find the
right scaling when this happens.

Returning to the general  problem, one option  is to keep both the
fermions and
bosons
together and not eliminate only  non-singular modes. This is what one
does in
problems where gauge fields couple to fermions. In such cases the
following question arises.  For bosons the energy is measured from
the
origin in momentum space while for fermions it is measured from the
FS.
How are high and low energy modes to be defined? A boson that imparts
a
momentum parallel to the fermions momentum can take it out of the
cut-off (and is a high energy boson) while the same boson, if it
attaches itself to a fermion with a perpendicular momentum, will move
it
along the FS and be a low energy boson. A common solution is to pick
a
point on the fermion's FS and treat bosonic momenta in radial and
angular directions distinctly. We find this approach to a
rotationally
invariant problem unsatisfactory , even though for reasons we do not
fully
understand, it might work. On the other hand  a recent paper by
Altshuler, Ioffe and Millis \cite{AIM} describes a class of problems
where such
a choice of
coordinates {\em is} warranted. These authors consider  a FS which is
not
rotationally invariant and only two small segments of which
(parallel under translation by a vector $\vec{Q}$) are important. In
this case there {\em is} a preferred fermion direction and using the
fermion
dispersion relation near these points to decide which bosons are high
energy and which are low energy seems legitimate. This is one very
promising
possibility   we intend to fully digest when  we resume
our RG program.   However, for isotropic problems involving bosons
and fermions, we believe there is still need for improvement. In the
meantime we can take the attitude that since no problem is really
isotropic, one may begin by attacking the gauge problems by starting
with, say an elliptical FS and trying the scheme of Altshuler {\em et
  al}.

\section{Conclusion}

We began with systems that have  quantum phase
transitions at zero coupling, which exist due to special symmetries
of
the FS. Two examples given in Reference
\cite{RMP} and discussed here were the planar superconductor (with a
time-reversal invariant spectrum) and planar spinless fermions at
half-filling (with nesting symmetry of the FS).  The goal was to
analyze systems with  transitions at nonzero, but small, coupling so
that a
perturbative RG would be possible. To this end we had to come up with
systems with such transitions, study them in mean-field to map out
the
phase structure, and then apply the RG to the transitions. The first
step
was accomplished by destroying the symmetries of the FS by adding a
term
$I$ to the hamiltonian. In the superconductor this was a parallel
magnetic field
and in the spinless fermion case it was a chemical potential, which
we saw could be accomplished once again by a parallel magnetic field.
The, second step, which led to the results emphasized here, is the

mean-field analysis which showed us the following :

\begin{itemize}

\item The two systems have a very similar three phase structure  and
phase  transitions between them. They have isomorphic formulas for
energy and susceptibilities. They naturally have similar windows for
the inhomogeneous state. Note that we do not expect the systems to
behave identically when fluctuations are considered,   since they
break different kinds of symmetries.

\item The phase at large $I$ (or small coupling) is a Fermi liquid,
which

  is  polarized  in the superconducting  case. The next phase as $I$
is
  reduced is has  a condensate that has a nonzero
  momentum {\em with respect to the canonical value}, this being zero
  for the superconductor and $(\pi ,\pi )$ for the CDW.  For this
reason
  we call it the  inhomogeneous phase. It has  gapless excitations
  along with the nonzero order parameter. We have computed some of
its
  properties for the superconductor in Section III. The last phase
(at small $I$) is a
  state with condensate at the canonical momentum, which is the BCS
state for the superconductor and the  $Q= (\pi , \pi  )$ state for
the CDW.

\item The transition from the Fermi liquid to the inhomogeneous state
is
  second order in mean-field but with nonclassical exponents: $\beta
=2$
  for the superconductor. The nonclassical exponents arise because
the
  chosen momentum for the condensate is a singular point of the Fermi
  liquid susceptibility.  The transition from the intermediate state
to
  the homogeneous  state is first order.

\item The large window for the inhomogeneous state is due to
  two-dimensionality. Indeed anisotropic systems have larger windows
  since they are more one-dimensional. Recall that for the CDW case
$$ {\Delta_0 \over \sqrt{2}}  < I < r \Delta_0 $$
 where $\Delta_0$ is the order parameter before any field is applied
and $r$ is the anisotropy. For the isotropic superconductor, set
$r=1$ above.

We  expect a similar enlargement of the window for anisotropic
superconductors.

\item In terms of experimental realizations, we require clean,
 truly two-dimensional, low $Z$ systems.

In the superconducting  case we need a  $B$ (in Tesla ) of order
$T_c$

(in Kelvin) assuming a free electron $g$ factor.

 In the CDW case, in an idealized world where
spinless fermions exist,  once again Zeeman energies of the order

of the gap will be needed.

In practice, to employ the trick discussed here,
 stronger fields may will be  needed to

separate real electrons into two sufficiently
 polarized species. Here a large $g$ factor will help.  \end{itemize}

Our analysis has many limitations. First it is mean-field. The order
of
the transitions may change. (The analysis of Altshuler {\em et al }
suggests both transitions may be first order.)  However, given the
large
windows, the intermediate phases should not disappear due to
fluctuations.

Our analysis ignores impurities and spin-orbit scattering. To see the
phase discussed here will require clean, low $Z$ two-dimensional
systems
and strong magnetic fields. For the CDW case, where we try to move
the  FS
using the field, a large $g$ factor will help. It appears that we are
at  the
threshold of
having all these.

Our analysis is at $T=0$. This restriction was not due to any
technical reasons and can be readily
overcome.

While our RG program has naturally led us to study this {\em family}
of
systems, the systems themselves are not new. As we mentioned,
Bulaevskii
has presaged many of our results  for the
superconductor. There is no doubt a similarly huge body of literature
on
the CDW problem, both theoretical and experimental.  Rather than
focus on
individual systems, we have    taken a slice through them  and
emphasized  the
common thread that runs through
many of them.  The next step is to go through individual cases that
have been
studied experimentally and see if they could belong to this family,
i..e, see
if they were clean enough, two-dimensional enough, satisfy the
various
assumptions made in the derivation  etc.  For example do our
consideration possibly apply to granular superconductors?
\cite{goldman}\cite{Adams} Do they apply to CDW transitions induced
by
application of pressure?\cite{pressure} We expect a careful analysis
of the huge literature (of which we have touched on a small sample
just to make our point) will  be
quite involved and
   welcome the  readers'  input.   In the meantime we  emphasize that
the

experimental detection and study of the

intermediate inhomogeneous phase in any one system  will not only be
fascinating in its   own right, but  also in terms of the family
structure emphasized  here.

{\bf Acknowledgements}\\

We thank  V. Ambegaokar, M.Beasely, J.Brooks, A. Goldman,
B.I.Halperin,
P.Hohenberg, K.Maki,
B.Maple, A.Millis, D.R.Nelson,  N.Read, H.Schulz, W.Skocpol, S.Sondhi
and
Y.Liu for helpful
discussions  on this subject, and are specially indebted to    Daniel
Prober and Subir Sachdev,  for  their generous time and assistance.
This
work was supported by  NSF Grants  DMR 9311949 and
DMR 9120525.

\newpage
\begin{center}
{ \bf{Figure Captions}}
\end{center}
Figure 1. Linear phase diagram for (a) the superconductor  and (b)
CDW systems
as $I$ is raised. In both cases, $I_0$ is where the Fermi liquid
susceptibility
for uniform condensate diverges. It does not correspond to a phase
transition.
In
the CDW case $I_q$ is when the  Fermi liquid susceptibility diverges
for
momentum is the $x$ direction.  This too is not a real transition
since  the
system has already condensed at $I_p$ to a state with momentum in the
$y$
direction.  All momenta are measured from $(\pi , \pi )$) in the CDW
case. \\
Figure 2. The dark lines show the anisotropic, two-branched, nested
Fermi
surface at
half-filling. The vector $\vec{Q}) $ connects any point on it to any
other
point. When the chemical potential is turned on, the surface shrinks
to the
thin line. There is no global nesting vector. \\
Figure 3. The energy of the  the inhomogeneous and uniform BCS states
relative
to the polarized Fermi liquid (PFL) as a function of the order
parameter
$\Delta$,  just when the BCS energy (labeled $q=0$ ) relative to
that of the
PFL.  vanishes. This is the point $I_F = \Delta_0 / \sqrt{2}$ in
Figure 1. Note
that that at this point the inhomogeneous state has a lower energy.
The optimal
momentum is somewhat larger than $I$. Given the energy scales, it
clear that
under a very slight reduction of $I$, the BCS state will dip below
the
inhomogeneous state.  Thus we may take $I_F = \Delta_0 / \sqrt{2}$ as
the
transition point between the uniform state and the inhomogeneous
state.\\
Figure 4. Phase structure as a function of number density $n$ as well
as
chemical potential $I$. As $I$ is raised, the system evolves along
the $x$
axis from the $(\pi , \pi )$ CDW  state to the inhomogeneous state
 to the PFL. As a function of $n$, slightest doping from half-filling
leads to
phase coexistence till $n$ comes down to $n_q$. Thereafter the
behavior is
quite smooth.

\end{document}